\documentclass[aps,prx,twocolumn,longbibliography]{revtex4-2}

\usepackage{amssymb}
\usepackage{amsbsy}
\usepackage{amsmath}
\usepackage{graphicx}
\usepackage{graphics}
\usepackage{setspace}
\usepackage{array}
\usepackage{color}
\usepackage{fontenc}
\usepackage{textcomp}
\usepackage{bm}
\usepackage{float}
\usepackage[bookmarks=false,linkcolor=blue,urlcolor=blue,colorlinks,citecolor=blue]{hyperref}

\newcommand{\vex}[1]{\bm{\mathrm{#1}}}

\newcommand{\bsub}{\begin{subequations}}
\newcommand{\esub}{\end{subequations}}

\graphicspath{{../Figures/}}

\begin{document}
\title{Multilayer graphene with a superlattice potential}
\author{Sayed Ali Akbar Ghorashi$^1$}\email{sayedaliakbar.ghorashi@stonybrook.edu}
\author{Jennifer Cano$^{1,2}$}
\affiliation{$^1$Department of Physics and Astronomy, Stony Brook University, Stony Brook, New York 11974, USA}
\affiliation{$^2$Center for Computational Quantum Physics, Flatiron Institute, New York, New York 10010, USA}

\date{\today}

\newcommand{\be}{\begin{equation}}
\newcommand{\ee}{\end{equation}}
\newcommand{\bea}{\begin{eqnarray}}
\newcommand{\eea}{\end{eqnarray}}
\newcommand{\h}{\hspace{0.30 cm}}
\newcommand{\vs}{\vspace{0.30 cm}}
\newcommand{\n}{\nonumber}

\begin{abstract}

Bernal stacked bilayer graphene subject to a superlattice potential can realize topological and stacked flat bands~\cite{GhorashiBLGSL}.
In the present work, we extend the study of a superlattice potential on graphene heterostructures to trilayer and quadrilayer graphene.
Comparing Bernal- and chirally-stacked multilayers reveals that the latter are more suitable for realizing stacks of many flat bands.
On the other hand, Bernal-stacked graphene heterostructures can realize topological flat bands.
Imposing two simultaneous superlattice potentials enhances the viability of both regimes.
\end{abstract}
\maketitle

\section{Introduction}
Twisted heterostructures have recently emerged as a highly tunable family of materials exhibiting a remarkable range of phenomena and showing promise as a quantum simulation platform for strongly correlated and topological physics \cite{bistritzer2011moire,cao2018unconventional,yankowitz2019tuning,lu2019superconductors,serlin2020intrinsic,nuckolls2020strongly,chen2020tunable,xie2021fractional,cao2018correlated,xu2020correlated,wang2021chiral,tang2020simulation,regan2020mott,Kennes2021}.
The most prominent example is twisted bilayer graphene (TBLG), which can host strongly correlated insulating phases and superconductivity \cite{cao2018correlated,cao2018unconventional}. Subsequent studies on twisted multilayer graphene (MLG) heterostructures also exhibit correlation-driven physics \cite{liu2020tunable,he2021symmetry,burg2019correlated, shen2020correlated, cao2020tunable1, park2021tunable,xu2021tunable, chen2021electrically, hao2021electric,park2021magic}.
Twisted bilayers beyond graphene have further demonstrated the versatility and tunability of these platforms \cite{regan2020mott,xu2020correlated,PhysRevLett.122.086402,wang2020correlated,PhysRevLett.121.026402,tang2020simulation,devakul2021magic,zang2021hartree,bi2021excitonic,wang2021staggered,zang2021dynamical,wietek2022tunable,xian2021realization,Klebl_2022,hejazi2020noncollinear,twistedbilayerCrI3,song2021direct,volkov2020magic,can2021high,zhao2021emergent,AaronJenTISL,guerci2022chiral}.

Despite the excitement and rapid progress, twisted heterostructures also suffer from new types of disorder, such as inhomogenous angle and strain, lattice relaxation, and sensitivity to the substrate. Together these factors severely impact sample reproducibility \cite{lau2022reproducibility}. Hence, alternative approaches to realize flat bands and moir\'e physics is desirable.

Conceptually, twisting adjacent layers induces a spatially modulated interlayer coupling that quenches the kinetic energy, resulting in flat bands and interaction-dominated physics.
We propose a change of paradigm where interlayer modulation is replaced by intralayer modulation, in the form of a superlattice potential.
A spatially modulated superlattice potential can be achieved by inserting a patterned dielectric between the gate and the sample, a set-up that has been realized experimentally~\cite{forsythe2018band,li2021anisotropic,barcons2022engineering}
and studied theoretically~\cite{park2008anisotropic,PhysRevLett.101.126804,park2009making,PhysRevB.81.075438,dubey2013tunable,ponomarenko2013cloning,forsythe2018band,huber2020gate,li2021anisotropic,lu2022synergistic} in monolayer graphene.
The effect of a superlattice potential on bilayer graphene (BLG) has also been studied \cite{BLGSLArunPRL,BLGSLArunPRB,ramires2018electrically}, but flat band physics was not considered.
Superlattice potentials applied to other systems can also promote correlation-driven physics~\cite{JenTISL,wang2021moire,JenDaniele,shi2019gate}.

Recently, we revisited BLG with a superlattice potential and showed that under experimentally feasible conditions, two distinct regimes of flat bands can be achieved, exhibiting either topological flat bands or a stack of many flat bands \cite{GhorashiBLGSL}.

In the present work, we ask under what conditions the two distinct regimes can be realized in MLG with a superlattice potential. In particular, we studied a superlattice potential on trilayer graphene (TLG) and quadrilayer graphene (QLG)
for both the case of chiral (ABC, ABCA) and Bernal (ABA, ABAB) stacked structures. We find that chirally stacked MLG favors the stacked flat band phase. On the other hand, Bernal TLG  and QLG are more suitable for topological flat bands, although the latter requires fine-tuning unless an additional sublattice mass is included.
In all cases, unlike in BLG, the remote hopping terms cannot be neglected.

The superlattice potential becomes less effective as more graphene layers are added due to screening.
This motivates us to consider a second set-up where MLG is sandwiched between two superlattice gates.
In this case, chiral stacked MLG can realize both flat band regimes.

\begin{figure}[htb!]
    \centering
    \includegraphics[width=0.5\textwidth]{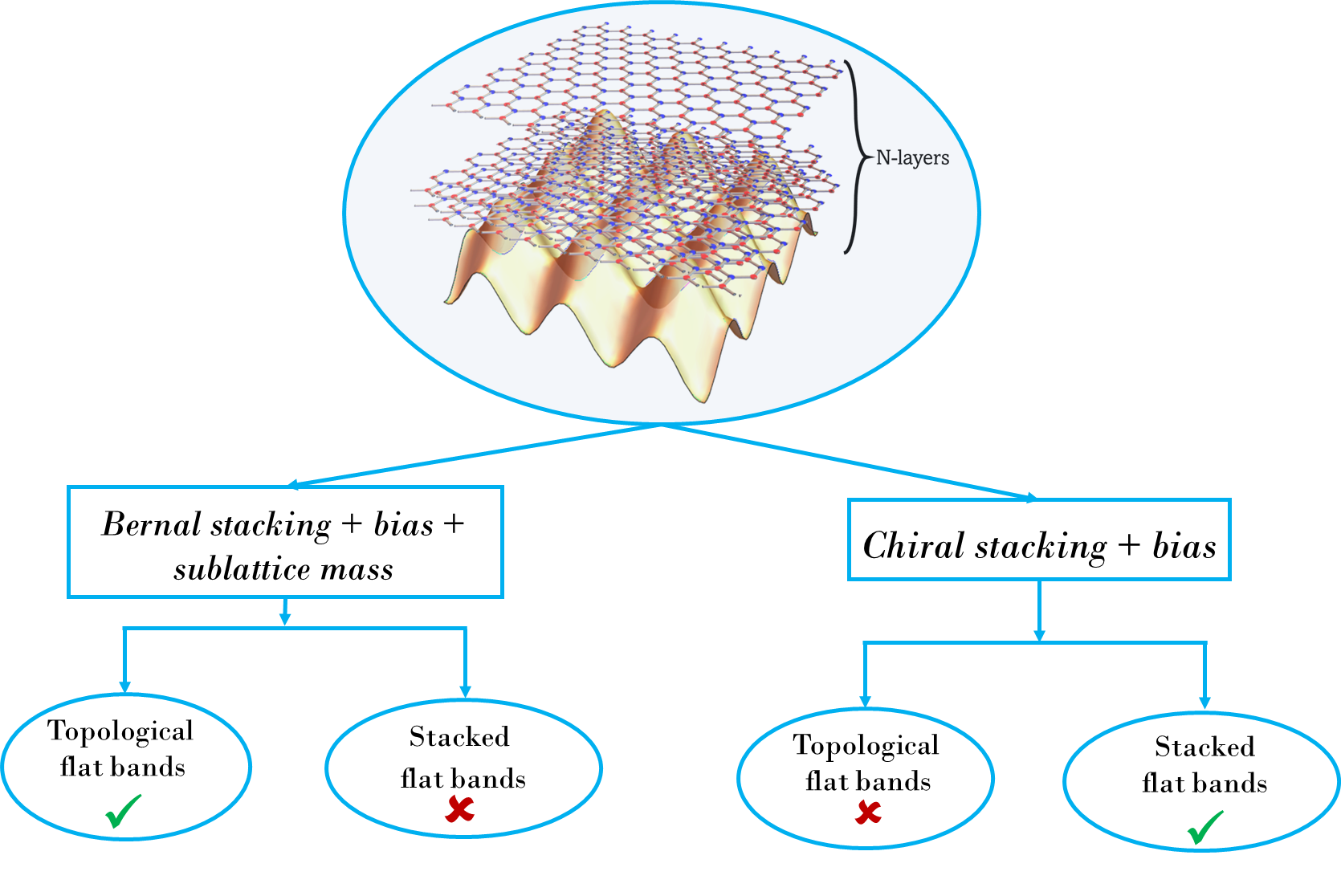}
    \caption{Summary of results. Biased chirally stacked MLG favors stacked flat bands while Bernal stacked MLG favors topological flat bands, assisted by a sublattice mass.}
    \label{fig:my_label}
\end{figure}
\begin{figure*}[htb!]
    \centering
    \includegraphics[width=1.05\textwidth]{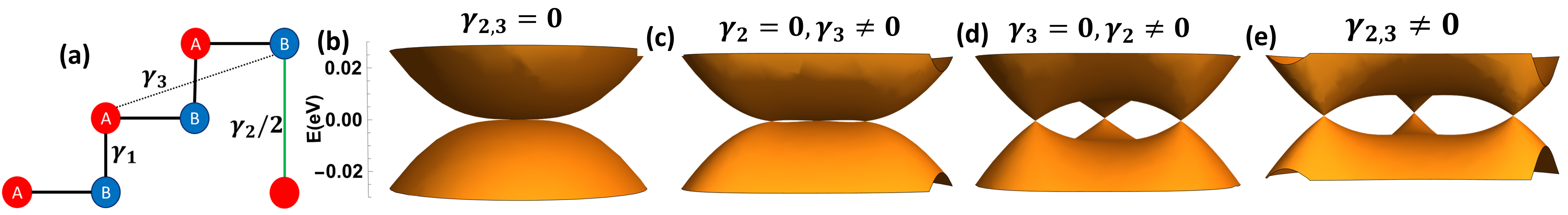}
    \caption{(a) Schematic of hopping processes in
our model of ABC-stacked TLG. (b-e) Band structures of ABC stacked TLG for different values of remote hoppings. We used $\gamma_1=0.4,\,\gamma_2=-0.02,\,\gamma_3=0.308$ eV.}
    \label{fig:ABC_vsl0}
\end{figure*}
\begin{figure}[hbt!]
    \raggedleft    \includegraphics[width=0.49\textwidth]{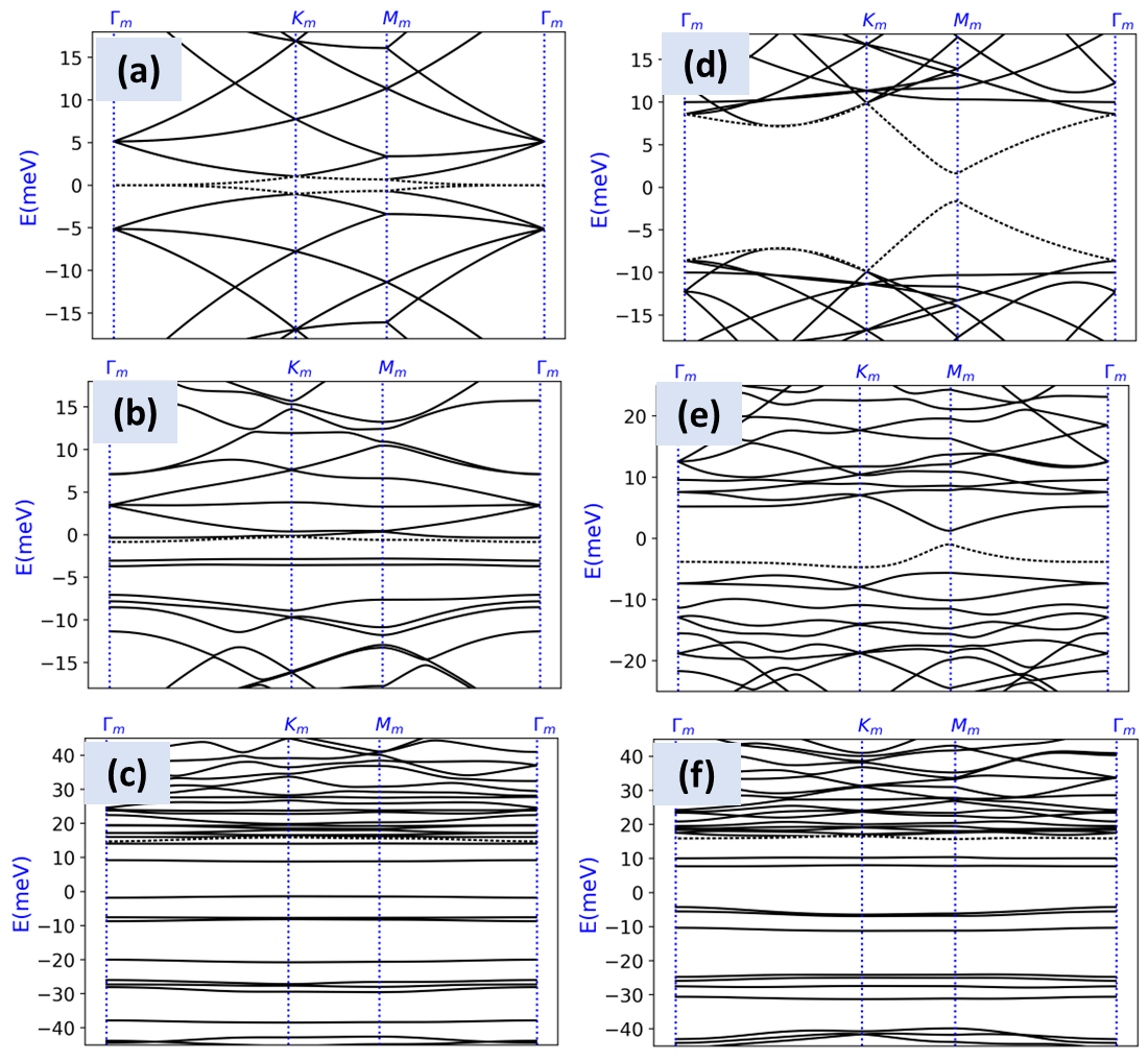}
    \caption{Band structures of ABC-stacked TLG with a triangular superlattice potential, without (a-c) and with (d-f) remote hoppings. (a,d) $V_{SL}=0,\,V_0=0$ meV, (b,e) $V_{SL}=5,\,V_0=0$ meV, (c,f)  $V_{SL}=20,\,V_0=-20$ meV. While the low-energy band structure for zero or weak $V_{SL}$ appears gapped in the presence of the remote hoppings (d,e), there are gapless points off the high-symmetry plotted path, consistent with Fig.~\ref{fig:ABC_vsl0}.}
    \label{fig:ABC_hex_vslv0}
\end{figure}

\section{Chiral stacking}

We start by discussing chirally stacked multilayer graphene where the the $A$ sites of the upper layer are above the $B$ sites of the lower layer and coupled by the nearest neighbor interlayer hopping amplitude, as shown in Fig.~\ref{fig:ABC_vsl0}(a). In the following we focus on two examples, ABC-stacked TLG and ABCA-stacked QLG to elucidate the effect of a superlattice potential on chirally stacked MLG.

\subsection{ABC}

We start with the brief review of ABC-stacked trilayer graphene in the absence of a superlattice potential. In the basis $(A_1,B_1,A_2,B_2,A_3,B_3)$, a low-energy Hamiltonian can be written as \cite{min2008electronic}
\begin{align}\label{HABC}
    H_{ABC}(\vex{k})=\left(\begin{array}{cccccc}
       0  & v \pi^{\dagger} & 0 & v_3\pi & 0 & \gamma_2/2\\
       v\pi & 0 & \gamma_1 & 0 & 0 & 0 \\
       0 & \gamma_1 & 0 & v\pi^{\dagger} & 0 & v_3\pi \\
       v_3\pi^{\dagger} & 0 & v\pi & 0 & \gamma_1 & 0 \\
       0 & 0 & 0 & \gamma_1 & 0 & v\pi^{\dagger} \\
       \gamma_2/2 & 0 & v_3\pi^{\dagger} & 0 & v\pi & 0
    \end{array}\right),
\end{align}
where $\pi=\chi k_x+ik_y$, $\chi=+(-)$ for valley $K(K')$, $v=(\sqrt{3}/2)a\gamma_0/\hbar$ denotes the Fermi velocity of each graphene layer, $\gamma_2$ parameterizes intrasublattice hopping between the top and bottom layers, and $v_3=(\sqrt{3}/2)a\gamma_3/\hbar$ accounts for trigonal warping.

Fig.~\ref{fig:ABC_vsl0} shows the band structure of \eqref{HABC} with and without the remote hoppings $\gamma_{2,3}$, which are depicted in Fig.~\ref{fig:ABC_vsl0}(a). In the absence of remote hoppings, i.e. $\gamma_{2,3}=0$, the low-energy bands have a cubic dispersion, as shown in Fig.~\ref{fig:ABC_vsl0}(b). The addition of $\gamma_{2,3}\neq 0$ splits the cubic Dirac fermion into three linearly dispersing cones, as shown in Figs.~\ref{fig:ABC_vsl0}(c)-(e). The remote hoppings are relevant within around $10-20$ meV of the charge neutrality point. In contrast, the analogous terms in bilayer graphene are only important within a few meV of the Dirac point \cite{PhysRevB.82.035409}.

The effect of a displacement field is included via
\begin{align}
    H_{V_0}=\left(\begin{array}{ccc}
       V_0\mathbb{I}_2  & \mathbf{0}_{2} & \mathbf{0}_{2} \\
      \mathbf{0}_{2}   & \mathbf{0}_{2} & \mathbf{0}_{2}\\
      \mathbf{0}_{2} & \mathbf{0}_{2} & -V_0\mathbb{I}_2 \\
    \end{array}\right),
    \label{V03}
\end{align}
which breaks inversion symmetry and opens up a gap in chirally stacked MLG.

We now consider the effect of a spatially modulated superlattice potential, described by
\begin{align}
    H_{SL}(\mathbf{r})=& \frac{V_{SL}}{2}\left(\begin{array}{ccc}
       \mathbb{I}_2  & \mathbf{0}_{2} & \mathbf{0}_{2} \\
      \mathbf{0}_{2}   & \alpha\mathbb{I}_2 & \mathbf{0}_{2}\\
      \mathbf{0}_{2} & \mathbf{0}_{2} & \beta\mathbb{I}_2 \\
    \end{array}\right)\sum_{n} \cos(\vex{Q}_n \cdot \bf{r}),
    \label{HSL3}
\end{align}
where $V_{SL}$ is the strength of the superlattice potential and the set of $\vex{Q}_n$ are its wave vectors.
We specialize to the case of a triangular superlattice potential with $\vex{Q}_n=Q(\cos(2n\pi/6),\sin(2n\pi/6))$, $n=1,2,3$, which define the ``mini Brillion zone'' (mBZ) by $\Gamma_m = (0,0)$, $M_m = \frac{1}{2}\vex{Q}_0$, and $K_m = \frac{1}{3}\left( \vex{Q}_0 + \vex{Q}_1\right)$.
Note that $\Gamma_m$ corresponds to the original $K$ point of TLG.
The parameters $0\leq\alpha,\beta<1$ are the ratio of the superlattice potential felt on one layer relative to the other; the asymmetry between the layers results from the experimental set-up where the superlattice potential is applied only on one side of the multilayer graphene. In the following, we use $\alpha =0.3$ and $\beta=0.1$ for both stackings of TLG  \cite{rokni2017layer}.

Fig.~\ref{fig:ABC_hex_vslv0} shows the evolution of the band structure of TLG in the triangular superlattice potential versus $V_{SL}$ and $V_0$, with (first column) and without (second column) remote hoppings. For comparison to our previous results for BLG~\cite{GhorashiBLGSL}, we took the superlattice period to be $L=50$ nm, although a smaller period may be preferred to produce more isolated flat bands.

Fig.~\ref{fig:ABC_hex_vslv0} shows similarities and differences compared to BLG \cite{GhorashiBLGSL}.
First, in BLG, isolated topological flat bands appeared in the limit of small $V_0$ and $V_{SL}$; we do not observe such bands in chiral TLG.
In the opposite limit where the fields are turned up, a regime of many stacked flat bands appears in both BLG and TLG.
In TLG, this regime appears at smaller field strength due to the larger density of states near the charge neutrality point.

We further observe that the remote hoppings cause the flat bands to develop a dispersion (Fig.~\ref{fig:ABC_hex_vslv0}(e)) compared to the case without remote hopping (Fig.~\ref{fig:ABC_hex_vslv0}(b)).
As the fields are increased, stacked flat bands form with and without remote hopping (Figs.~\ref{fig:ABC_hex_vslv0}(c,f)).

\begin{figure}[htb!]
    \centering
    \includegraphics[width=0.4\textwidth]{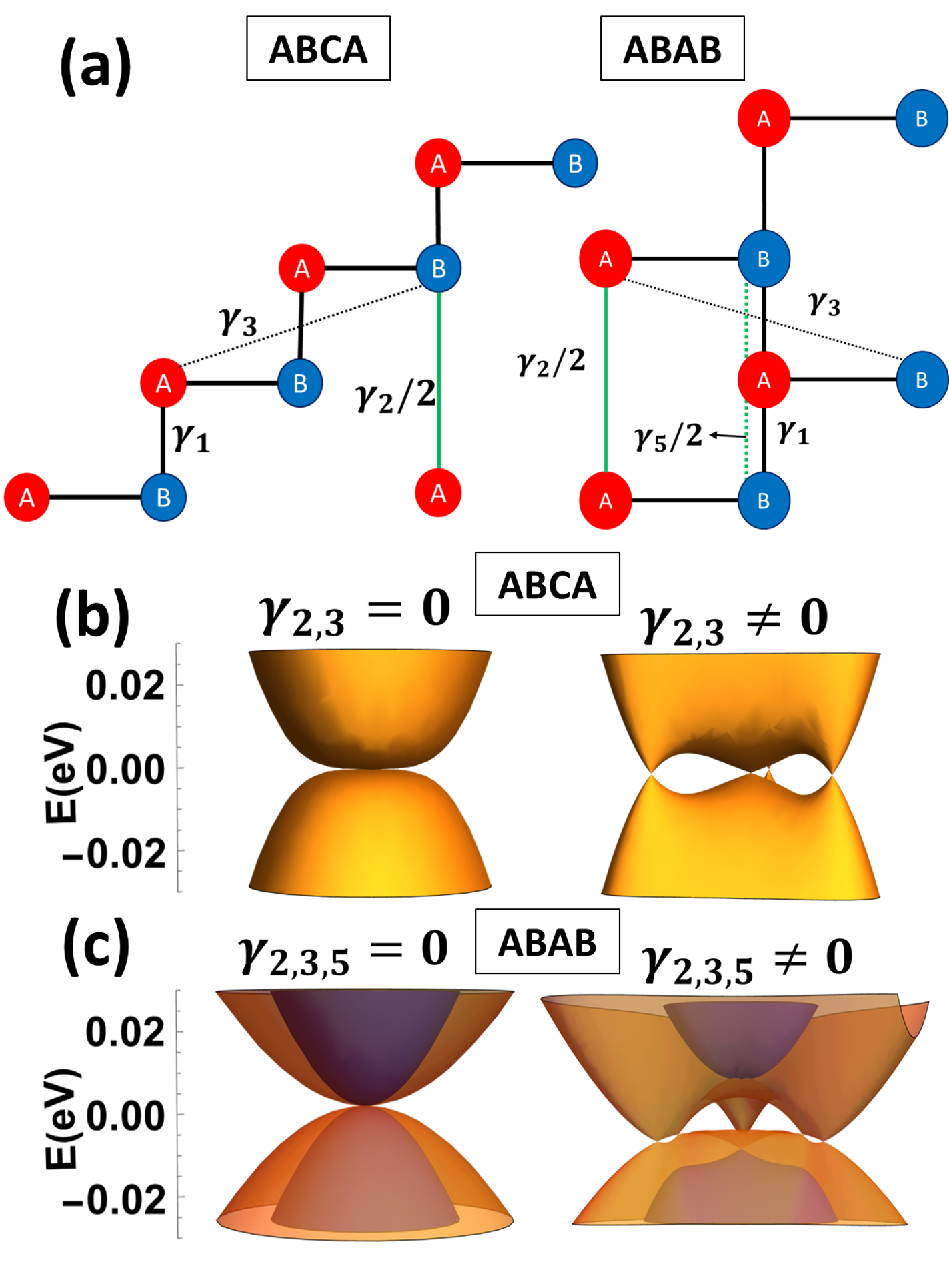}
    \caption{(a) Schematic representation of the hopping processes included in
our model for ABCA and ABAB stacked QLG. Band structure of (b) ABCA-stacked and (c) ABAB stacked QLG with and without the remote hoppings. $\gamma_1=0.4,\,\gamma_2=-0.02,\,\gamma_3=0.308,\,\gamma_5=0.04$ eV.}
    \label{fig:QLG_Vsl0}
\end{figure}

\subsection{ABCA}
ABCA-stacked QLG is a chiral multilayer graphene structure with four layers, as shown in Fig.~\ref{fig:QLG_Vsl0}(a). It can be described by the following Hamiltonian \cite{min2008electronic}
\begin{align}
    H_{ABCA}(\vex{k})=\left(\begin{array}{cccc}
    H_{\vex{G}} & \Gamma_1 & \Gamma_2 & \mathbf{0}_{2} \\
    \Gamma_1^{\dagger} & H_{\vex{G}} & \Gamma_1 & \Gamma_2 \\
    \Gamma_2^{\dagger} & \Gamma_1^{\dagger} & H_{\vex{G}} & \Gamma_1 \\
    \mathbf{0}_2 & \Gamma_2^{\dagger} & \Gamma_1^{\dagger} & H_{\vex{G}}
    \end{array}\right),
\end{align}
where $H_{\vex{G}}$ is the Dirac cone Hamiltonian in each layer and the interlayer couplings $\Gamma_{1,2}$ are given by
\begin{align}
    \Gamma_1=\left(\begin{array}{cc}
       0  & v_3\pi^{\dagger}\\
       \gamma_1  & 0
    \end{array}\right), \quad \Gamma_2=\left(\begin{array}{cc}
       0  & \gamma_2/2 \\
       0  & 0
    \end{array}\right)
\end{align}
Similar to TLG, we define the displacement field and superlattice potentials by
\begin{align}
  H_{V_0}=V_0\left(\begin{array}{cccc}
       \mathbb{I}_2  & \mathbf{0}_{2} & \mathbf{0}_{2} & \mathbf{0}_{2} \\
      \mathbf{0}_{2}   & \mathbf{0}_{2} & \mathbf{0}_{2} & \mathbf{0}_{2}\\
      \mathbf{0}_{2} & \mathbf{0}_{2} & \mathbf{0}_{2} & \mathbf{0}_{2}\\
      \mathbf{0}_{2} & \mathbf{0}_{2} & \mathbf{0}_{2} & -\mathbb{I}_2
    \end{array}\right)
\end{align}
and
\begin{align}
    H_{SL}(\mathbf{r})=& \frac{V_{SL}}{2}\left(\begin{array}{cccc}
      \mathbb{I}_2  & \mathbf{0}_{2} & \mathbf{0}_{2} & \mathbf{0}_{2} \\
      \mathbf{0}_{2}   & \alpha\mathbb{I}_2 & \mathbf{0}_{2} & \mathbf{0}_{2}\\
      \mathbf{0}_{2} & \mathbf{0}_{2} & \beta\mathbb{I}_2 & \mathbf{0}_{2}\\
      \mathbf{0}_{2} & \mathbf{0}_{2} & \mathbf{0}_{2} & \eta\mathbb{I}_2
    \end{array}\right)\sum_{n} \cos(\vex{Q}_n \cdot \bf{r}),
    \label{HSL4}
\end{align}
where we use $\alpha=0.3,\,\beta=0.1,\,\eta=0$ throughout this work \cite{rokni2017layer}. In the absence of remote hoppings the low-energy bands have a quartic dispersion. The remote hoppings split the quartic band touching point into the four linearly dispersing cones, shown in Fig.~\ref{fig:QLG_Vsl0}(b).
\begin{figure}[htb!]
    \centering
    \includegraphics[width=0.49\textwidth]{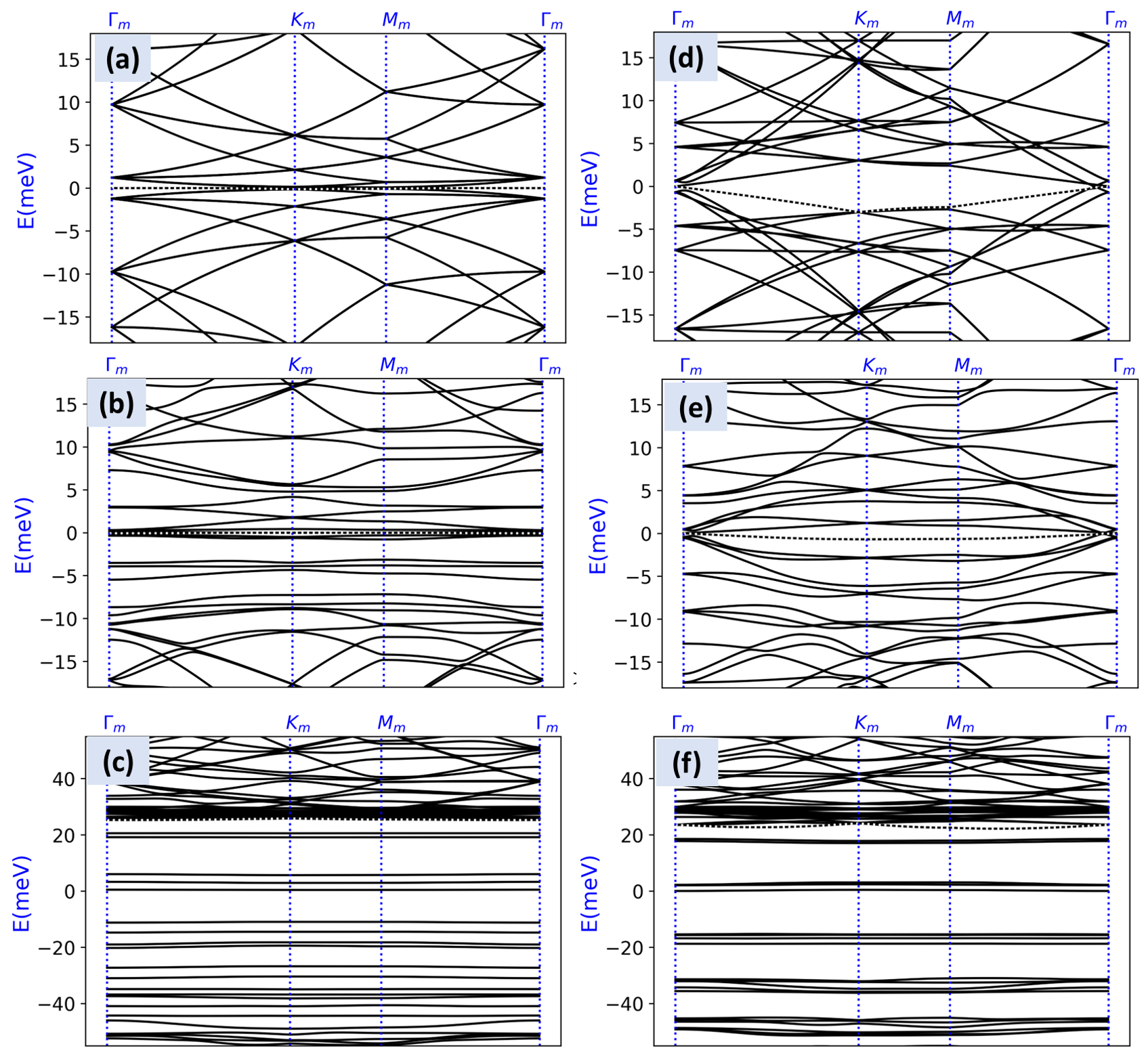}
    \caption{The band structures of ABCA-stacked QLG without (a-c) and with (d-f) remote hoppings. (a,d) $V_{SL}=0,\,V_0=0$ meV, (b,e) $V_{SL}=5,\,V_0=0$ meV, (c,f) $V_{SL}=20,\,V_0=-30$ meV.}
    \label{fig:ABCA_HEX_VslV0}
\end{figure}
\begin{figure*}[htb!]
    \centering
    \includegraphics[width=1.0\textwidth]{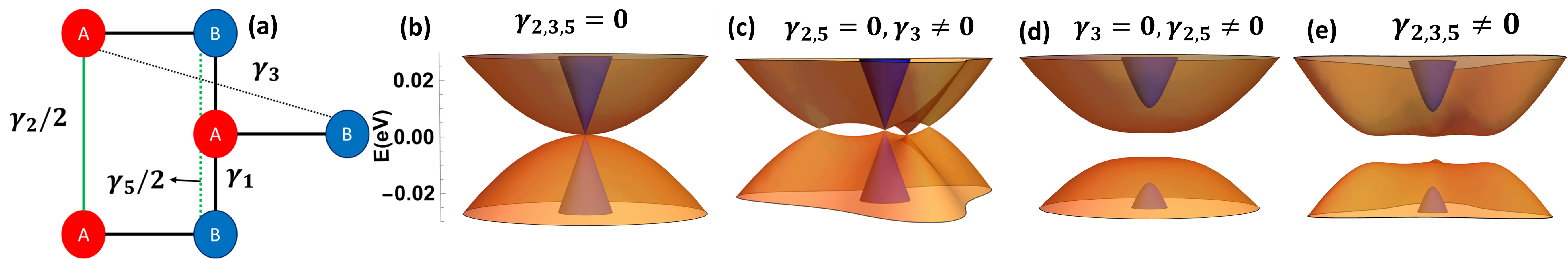}
    \caption{(a) Schematic representation of the hopping processes included in
our model for ABA-stacked TLG. (b-e) band structures of ABA stacked TLG for various configuration of remote hoppings. $\gamma_1=0.4,\,\gamma_2=-0.02,\,\gamma_3=0.308,\,\gamma_5=0.04$ eV}
    \label{fig:ABA_vsl0}
\end{figure*}

Fig.~\ref{fig:ABCA_HEX_VslV0} shows band structures of ABCA QLG in a superlattice potential.
Similar to chiral TLG, for small fields, the bands near charge neutrality flatten but do not become isolated.
As the fields are turned up, the stacked flat band regime appears;
comparing QLG, TLG and BLG, we observed a trend that for more layers of graphene, the stacked flat band regime appears for smaller fields, due to the larger density of states near the charge neutrality point.
Comparing Figs.~\ref{fig:ABCA_HEX_VslV0}(c,f) to Figs.~\ref{fig:ABC_hex_vslv0}(c,f), the remote hopping terms have a more prounounced effect for QLG than for TLG, causing the flat bands to clump together.

\section{Bernal stacking Graphene}

We now turn to studying the effect of a superlattice potential on Bernal-stacked MLG, specifically ABA TLG and ABAB QLG for three and four layers, respectively.

\subsection{ABA}

ABA-stacked TLG is shown in Fig.~\ref{fig:ABA_vsl0}(a).
Its Hamiltonian can be written as follows,
\begin{align}\label{HABA}
    H_{ABA}(\vex{k})=\left(\begin{array}{cccccc}
       0  & v \pi^{\dagger} & 0 & v_3\pi & \gamma_2/2 & 0\\
       v\pi & 0 & \gamma_1 & 0 & 0 & \gamma_5/2 \\
       0 & \gamma_1 & 0 & v\pi^{\dagger} & 0 & \gamma_1 \\
       v_3\pi^{\dagger} & 0 & v\pi & 0 & v_3\pi^{\dagger} & 0 \\
       \gamma_2/2 & 0 & 0 & v_3\pi & 0 & v\pi^{\dagger} \\
      0 & \gamma_5/2 & \gamma_1 & 0 & v\pi & 0
    \end{array}\right)
\end{align}
In ABA-stacked TLG the AA ($\gamma_2$, solid green bond in Fig.~\ref{fig:ABA_vsl0}(a)) and BB ($\gamma_5$, dotted green bond in Fig.~\ref{fig:ABA_vsl0}(a)) remote hoppings between the first and third layers have different strength and sign. We use $\gamma_2=-0.02$ (same as ABC-stacked) and $\gamma_{5}=0.1, \gamma_1=0.04$ throughout this work \cite{PhysRevLett.121.167601}. The displacement field and superlattice potential are implemented by the same matrices as for ABC-stacked TLG, i.e., Eqs.~(\ref{V03}) and~(\ref{HSL3}).
In the absence of remote hopping terms, the spectrum of ABA-stacked TLG consists of a linear and a quadratic crossing, shown in Fig.~\ref{fig:ABA_vsl0}(b).
However, unlike ABC-stacked TLG, in Bernal-stacked ABA TLG the external displacement field $V_0$ does not gap out the spectrum completely.
Instead, the effect of $V_{0}$ is to gap the linear cone but leave the quadratic touching gapless.
When the remote hopping terms are included, the spectrum becomes fully gapped; turning up $V_0$ further increases the gap of the linear cone, but does not change the overall band gap.
As we will see, this has important consequences for the realization of the stacked band regime.

\begin{figure*}[htb!]
    \centering
    \includegraphics[width=1\textwidth]{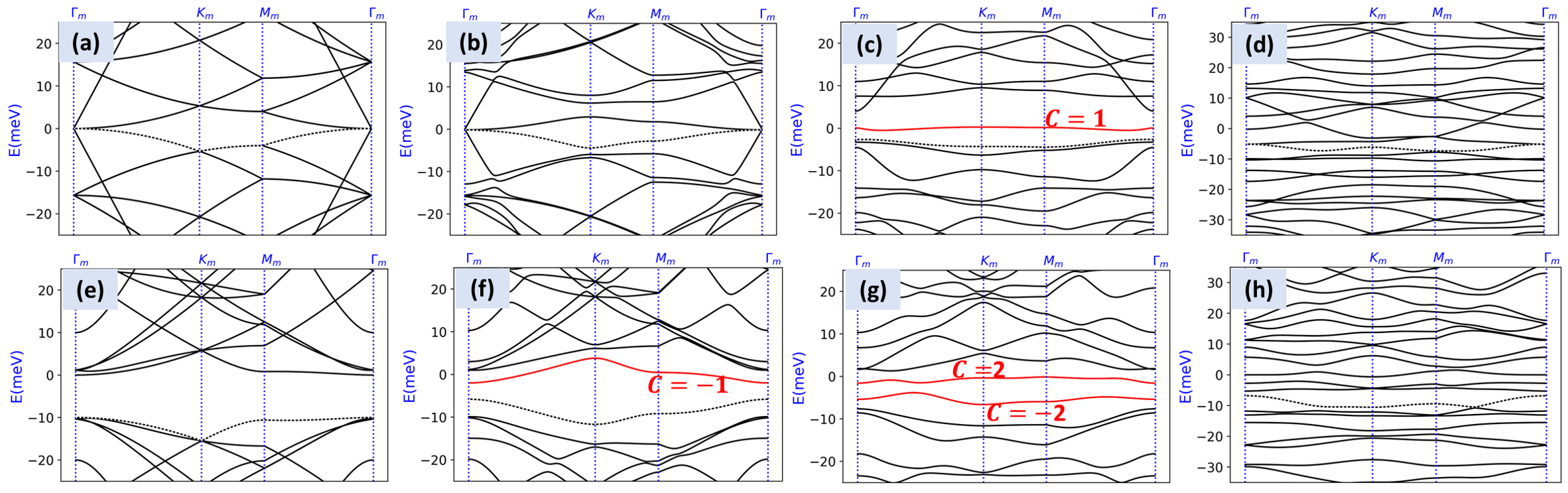}
    \caption{Band structure of ABA-stacked TLG with a triangular superlattice potential in absence of sublattice mass $m=0$ without (a-d) and with (e-h) remote hoppings. (a,e) $V_{SL}=0,\,V_0=0$ meV, (b,f) $V_{SL}=5,\,V_0=0$ meV, (c,g) $V_{SL}=17,\,V_0=0$ meV, (d,h) $V_{SL}=30,\,V_0=-60$ meV}
    \label{fig:ABA_VslV0}
\end{figure*}

Fig.~\ref{fig:ABA_VslV0} shows the effect of a superlattice potential and displacement field with $m=0$ on ABA-stacked TLG without (Fig.~\ref{fig:ABA_VslV0}(a-d)) and with (Fig.~\ref{fig:ABA_VslV0}(e-h)) the remote hopping terms.
Starting with $V_0=0$ and a weak superlattice potential, including the remote hoppings causes two bands to completely detach from the rest of the bands, as shown in Fig.~\ref{fig:ABA_VslV0}(f).
Importantly, one of the detached bands is topological with Chern number $C=-1$.
As $V_{SL}$ is further turned up, topological flat bands with higher Chern numbers can also emerge, as shown in  Fig.~\ref{fig:ABA_VslV0}(g).
However, even for a strong potential and non-zero displacement field the stacked flat band regime does not appear.

\begin{figure}[htb!]
    \centering
    \includegraphics[width=0.49\textwidth]{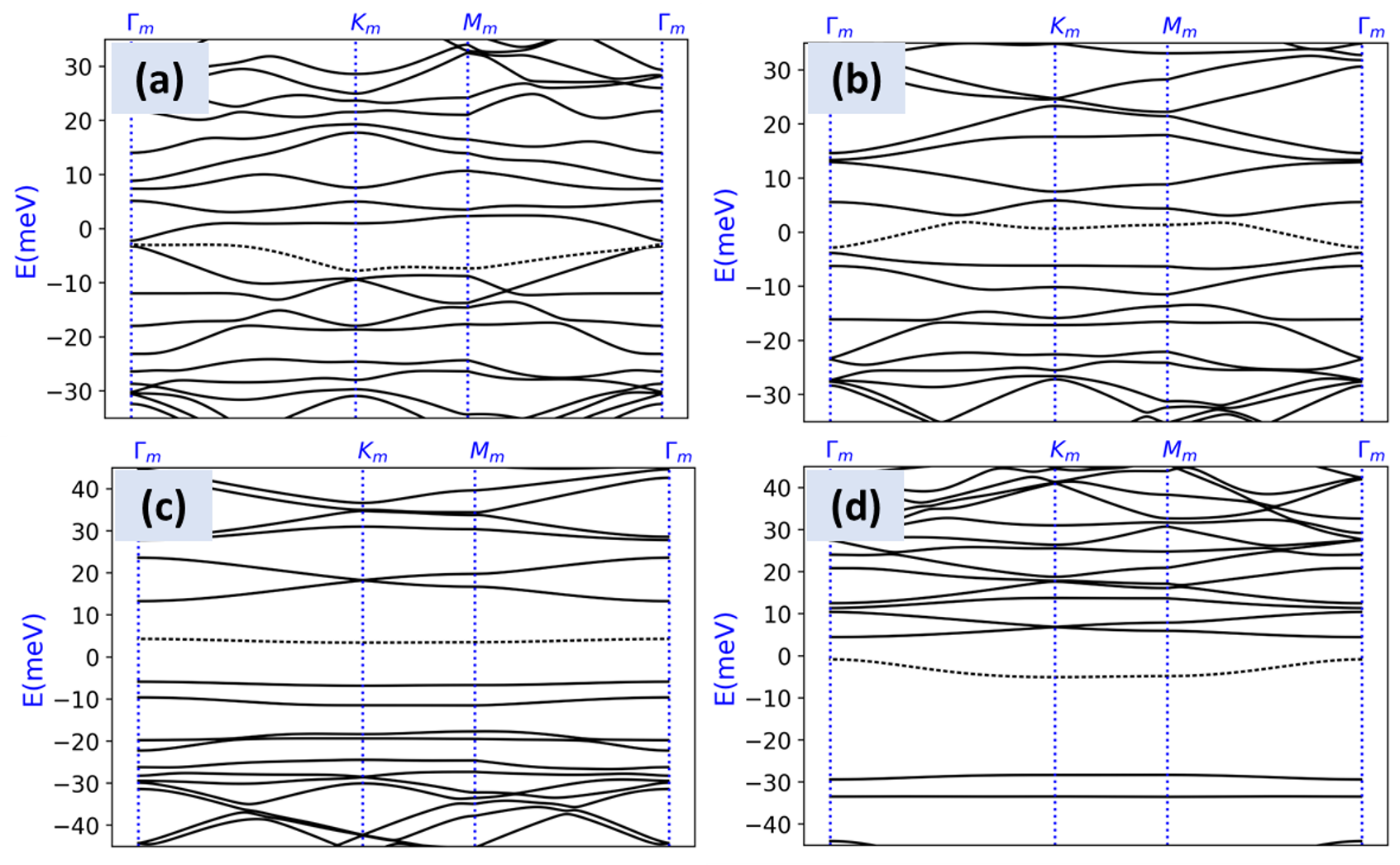}
    \caption{Band structure of ABA-stacked TLG,  varying the sublattice mass $m$ at fixed gate potentials $V_{SL}=20,\,V_0=0$ meV.
    (a-d) Spectrum with $m=10,\,30,\,50,\,-50$ meV, respectively.
    Stacks of flat bands start to appear in (c) and (d).
    In all plots, remote hopping terms are included and $\delta=0.3$.}
    \label{fig:ABA_Vslm}
\end{figure}

We hypothesize that the stacked flat band regime requires a large and tunable band gap in the $V_{SL}=0$ limit; intuitively, the large band gap provides an energy window that can host bands detached by the superlattice potential.
As explained below Eq.~(\ref{HABA}), $V_0$ does not change the low-energy band gap.
Thus, to test our hypothesis, we introduce a tunable band gap via a sublattice mass $m$ as follows:
\begin{align}
     H_{m}=m\left(\begin{array}{ccc}
       \sigma^z  & \mathbf{0}_{2} & \mathbf{0}_{2} \\
      \mathbf{0}_{2}   & \delta\sigma^z & \mathbf{0}_{2}\\
      \mathbf{0}_{2} & \mathbf{0}_{2} & \sigma^z \\
    \end{array}\right),
    \label{Hmass}
\end{align}
where $\sigma^{i}$ are Pauli matrices of sublattice space. The mass $m$ may result from TLG sandwiched between two layers of aligned hexagonal boron nitride, so that the middle layer feels only a ratio $0\leq\delta<1$ of mass $m$.
Although this set-up would not result in a tunable value of $m$, we treat $m$ as a tunable paramter to illustrate the vital effect of a tunable band gap. Fig.~\ref{fig:ABA_Vslm} shows that as $m$ is turned up, the stacked flat band regime appears, even with $V_0 = 0$.
However, compared to TLG, the stack of flat bands does not span as wide an energy range as in chiral TLG.

In summary, the phenomenology of Bernal TLG differs from chiral stacking in two ways: first, unlike chiral stacked TLG, Bernal stacked TLG can easily realize topological flat bands in the weak field limit. Second, the displacement field cannot induce the stacked flat band regime in Bernal stacked TLG; though the sublattice mass can play a similar role, it has less potency.

\subsection{ABAB}

We now discuss Bernal (ABAB-stacked) QLG, shown in Fig.~\ref{fig:QLG_Vsl0}(a), which is described by the following low-energy Hamiltonian,
\begin{align}\label{bernalQLG}
    H_{ABAB}(\vex{k})=\left(\begin{array}{cccc}
    H_{\vex{G}} & \Gamma_1 & \Gamma_2 & \mathbf{0}_{2} \\
    \Gamma_1^{\dagger} & H_{\vex{G}} & \Gamma_1^{\dagger} & \Gamma'_2 \\
    \Gamma_2^{\dagger} & \Gamma_1 & H_{\vex{G}} & \Gamma_1 \\
    \mathbf{0}_2 & \Gamma_2^{'\dagger} & \Gamma_1^{\dagger} & H_{\vex{G}}
    \end{array}\right),
\end{align}
where
\begin{equation}
    \Gamma_1=\left(\!\! \begin{array}{cc}
       0  & v_3\pi\\
       \gamma_1  & 0
    \end{array} \!\! \right)\!,
    \Gamma_2=\left( \!\! \begin{array}{cc}
       \gamma_2/2  & 0 \\
       0  & \gamma_5/2
    \end{array} \! \!\right)\!,
    \Gamma'_2=\left( \!\! \begin{array}{cc}
       \gamma_5/2  & 0 \\
       0  & \gamma_2/2
    \end{array} \!\! \right)
\end{equation}

The energy spectrum of Eq.~\eqref{bernalQLG} shows two pairs of quadratically dispersing bands in the absence of remote hoppings; see \ref{fig:QLG_Vsl0}(c).
Unlike Bernal TLG, the remote hopping terms do not gap the spectrum, but split the two quadratic nodes into three pairs of linear Dirac cones.
A new feature is that one of the nodes in each pair has a small energy shift and tilt with respect to other nodes.

\begin{figure}[htb!]
    \centering
    \includegraphics[width=0.49\textwidth]{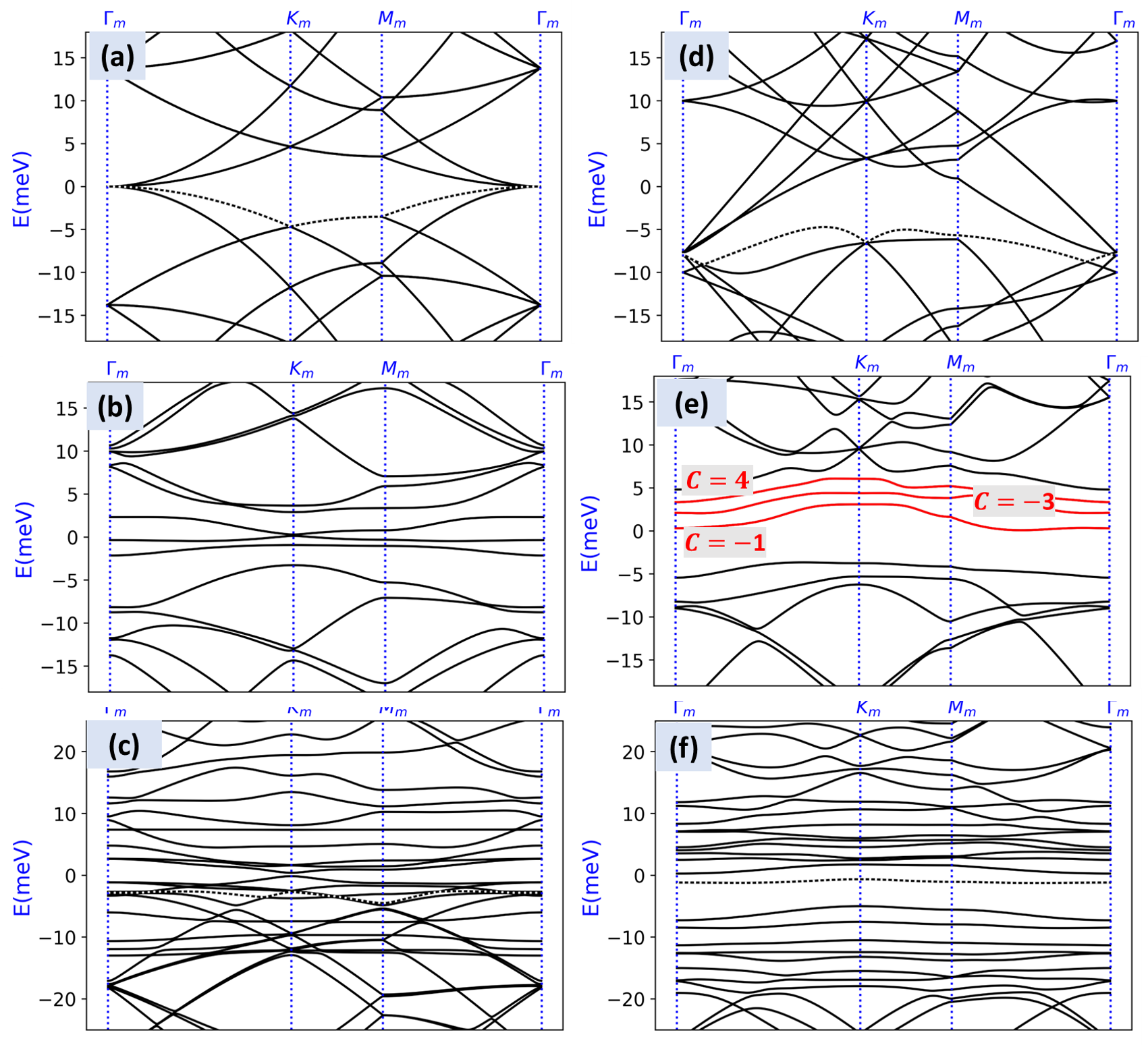}
    \caption{Band structure of ABAB-stacked QLG with a superlattice potential, without (a-e) and with (f-j) the remote hoppings; (a,d) $V_{SL}=0,\,V_0=0$ meV, (b,e) $V_{SL}=10,\,V_0=30$ meV, (c,f) $V_{SL}=20,\,V_0=-100$ meV.
    (b,e) For smaller potentials, topological flat bands can be realized when the remote hopping terms are included, although they are not isolated and require fine-tuning.
    For larger potentials, flat bands appear with and without remote hopping terms, as in (c) and (f), though they are not isolated.
    }
    \label{fig:ABA_HEX_VslsV0}
\end{figure}

The band spectra in the presence of the superlattice potential are shown in Fig.~\ref{fig:ABA_HEX_VslsV0}.
Fig.~\ref{fig:ABA_HEX_VslsV0}(d) shows the energy shift of the Dirac points when the remote hopping terms are included, as well as severe particle-hole symmetry breaking compared to (a).
In general, topological flat bands do not appear without fine-tuning, although they can appear when remote hopping terms are included, as in Fig.~\ref{fig:ABA_HEX_VslsV0}(e).
Increasing $V_0$ produces flatter bands both with and without the remote hopping terms, as shown in (c) and (f), although they are not isolated.
Thus, the stacked flat band regime is more accessible than in Bernal TLG but less favorable compared to the chiral-stacked structures.


Though isolated flat bands are not readily available in Bernal QLG, topological bands (including those with $|C|>1$) can be realized more generically when a sublattice mass term is included in addition to the displacement field, as shown in Fig.~\ref{fig:multiSL}(a) for a smaller value of $L$.
We incorporate the effect of a sublattice mass through the term
\begin{align}
     H_{m}=m\left(\begin{array}{cccc}
       \sigma^z  & \mathbf{0}_{2} & \mathbf{0}_{2} & \mathbf{0}_{2} \\
      \mathbf{0}_{2}   & \delta_1\sigma^z & \mathbf{0}_{2} & \mathbf{0}_{2}\\
      \mathbf{0}_{2} & \mathbf{0}_{2} & \delta_2\sigma^z & \mathbf{0}_{2} \\
      \mathbf{0}_{2} & \mathbf{0}_{2} & \mathbf{0}_{2} & \sigma^z \\
    \end{array}\right),
    \label{Hmass4}
\end{align}
and use $\delta_1=\delta_2=0.3$ throughout.

The smaller value of $L$ is important to achieve isolated flat bands in multilayer graphene.
Until now, we have fixed the superlattice periodicity at $L=50$ nm for comparison to Ref.~\onlinecite{GhorashiBLGSL}.
However, as the number of layers increases, the low-energy density of states also increases.
The increased density of states decreases the inter-band spacing in the presence of a superlattice potential, making the topological indices of individual bands sensitive to small parameter changes.
Further, experimentally, closely spaced bands will be difficult to resolve in the presence of disorder.
Thus, as the number of layers is increased, a smaller value of $L$ is preferred to achieve isolated flat bands.
Additional band structures with smaller values of $L$ are shown in Fig.~\ref{fig:multiSL}.


\section{Two superlattice potentials}

We briefly discuss the possibility of employing two superlattice potentials, one on the bottom and the other at the top of the sample. The motivation for considering such a setup is that the effect of the superlattice potential becomes vanishingly small for the upper layers of multilayer graphene.

To model this set-up, the parameters $\alpha$, $\beta$, and $\eta$, which parameterize the superlattice potential in each layer relative to the first layer, as defined in Eqs.~(\ref{HSL3}) and (\ref{HSL4}), must be modified. To reflect the symmetry of the set-up, we choose $\alpha = 0.5$, $\beta = 1$ in TLG and $\alpha=0.3$, $\beta=0.3$, and $\eta=1$ in QLG.

\begin{figure}[htb!]
    \centering
    \includegraphics[width=0.5\textwidth]{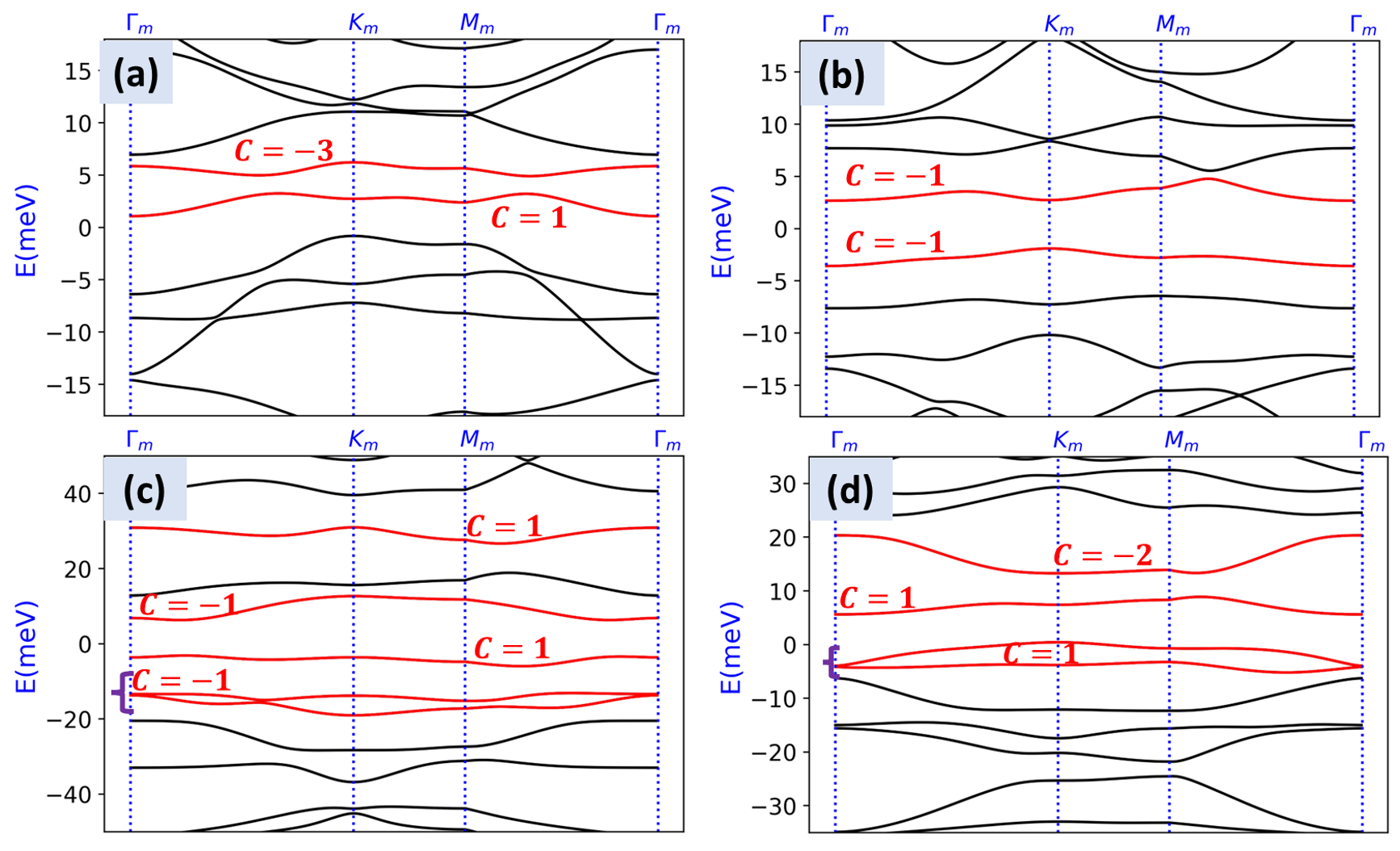}
    \caption{Two superlattice potentials realize topological flat bands (red) in MLG. (a) Bernal QLG with a single superlattice potential with $V_{SL}=40,\,V_0=60,\,m=20$ meV, $L=40$ nm, $\alpha=0.3,\,\beta=0.1,\,\eta=0$,  and $\delta_1=\delta_2=0.3$; (b) Bernal QLG with two superlattice potentials, $V_{SL}=20,\,V_0=50,\,m=20$ meV, $L=40$ nm, $\alpha=0.3,\,\beta=0.3,\,\eta=1$,  and $\delta_1=\delta_2=0.3$; (c) chiral QLG with with two superlattice potentials, $V_{SL}=25,\,V_0=20$ meV, $L=25$ nm and $\alpha=0.3,\,\beta=0.3,\,\eta=1$; (d) chiral TLG with with two superlattice potential $V_{SL}=30,\,V_0=30$ meV, $L=35$ nm and $\alpha=0.5,\,\beta=1$. Parameters are chosen to minimize the bandwidth of topological bands.}
    \label{fig:multiSL}
\end{figure}

In Fig.~\ref{fig:multiSL}(b), we show the effect of two superlattice potentials on Bernal-stacked QLG. We find that the two superlattice setup is capable of generating flatter topological bands even for weaker fields compared to the single superlattice potential setup (Fig.~\ref{fig:multiSL}(a)) because more layers feel the potential.

In the chiral-stacked case with two potentials, new phenomena appear. As shown in Fig.~\ref{fig:multiSL}(c,d), for both chiral TLG and QLG, multiple topological bands with small bandwidth can appear, including those with $|C|>1$; the parameters in each figure are chosen to produce the flattest topological bands.
In contrast, topological flat bands never occurred in the single potential setups for TLG and QLG shown in Figs.~\ref{fig:ABC_hex_vslv0} and \ref{fig:ABCA_HEX_VslV0}.
We conclude that the two superlattice potential setup may offer advantages over a single superlattice potential in realizing topological flat bands.

\section{Concluding remarks and practical recipes for experiment}

We performed a comprehensive study of flat bands in multilayer graphene subject to a superlattice potential. In \cite{GhorashiBLGSL} we showed that a superlattice potential induces two distinct regimes of flat bands in bilayer graphene, namely topological and stacked flat bands. In this work, we discussed the realization of these phases in graphene structures with more layers, which can be either chirally or Bernal stacked.
Our conclusions provide a cookbook for experimental realizations of flat bands in MLG structures with a superlattice potential.

Our main conclusions are as follows:
(1) Chirally stacked graphene multilayers are ideal to realize the stacked flat band regime. The higher the number of layers, the easier it is to get to that regime.
(2) Bernal stacked graphene is favorable for realizing topological flat bands in the setups with a single superlattice potential.
It does not realize the stacked flat band regime and becomes worse by increasing the number of layers.
(3) A sublattice mass (e.g., provided by hBN) can induce a gap in Bernal stacked multilayers, which can help generate stacked flat bands. However, we do not have a physical system in mind where such a mass is tunable. Artificial metamaterials may be helpful to realize this tunable mass.
(4)  The remote hoppings impact the realization of isolated flat bands in MLG at small fields and become more relevant as the number of layers increases.
(5) Due to the increased low-energy density of states as the number of layers increases, a smaller superlattice period is imperative to obtain isolated flat bands.
(6) Using two superlattice potentials enhances the effect of the superlattice and may be promising to achieve both the topological and stacked flat bands in chirally stacked MLG.

Our results demonstrate the generality and robustness of both the topological and stacked regimes of flat bands in multilayer graphene with an artificial superlattice potential. Thus, these systems provide a promising platform for quantum simulation and an alternative to twisted heterostructures. Future theoretical and experimental studies will demonstrate the correlated phenomena that derive from both flat band regimes.

\section*{Acknowledgements}
This work was supported by the Air Force Office of Scientific Research under Grant No. FA9550-20-1-0260. J.C. is partially supported by the Alfred P. Sloan Foundation through a Sloan Research Fellowship. The Flatiron Institute is a division of the Simons Foundation.


%

\end{document}